\documentclass[prx,10pt,twocolumn,reprint,showkeys,superscriptaddress,nofootbib, longbibliography]{revtex4-1}

\usepackage{amsmath}    
\usepackage{graphicx}   
\usepackage{verbatim}   
\usepackage{color}      
\usepackage{subfigure}  
\usepackage{hyperref}   
\usepackage{SIunits}     

\usepackage{amsmath}
\usepackage{bm}
\usepackage{bbold}
\usepackage{float}
\usepackage{xfrac}

\newcommand{\sm}[0]{\sigma_{-}}
\renewcommand{\sp}[0]{\sigma_{+}}
\newcommand{\sx}[0]{\sigma_{x}}
\newcommand{\sxy}[0]{\sigma_{x,y}}
\newcommand{\sy}[0]{\sigma_{y}}
\newcommand{\sz}[0]{\sigma_{z}}

\newcommand{\gm}[0]{\gamma_{-}}
\newcommand{\gp}[0]{\gamma_{+}}
\newcommand{\gN}[0]{\gamma_{N}}
\newcommand{\gM}[0]{\gamma_{M}}
\newcommand{\lr}[0]{\lambda_{0}}
\newcommand{\lc}[0]{\lambda_{1}}
\newcommand{\ld}[0]{\lambda_{2}}
\newcommand{\gNM}[0]{\gamma_{NM}}
\newcommand{\bout}[0]{b_{\mathrm{out}}}

\newcommand{\vv}[0]{\mathbf{v}}
\newcommand{\drm}[0]{\mathrm{d}}
\newcommand{\erm}[0]{\mathrm{e}}
\newcommand{\Lm}[0]{\mathcal{L}}
\newcommand{\id}[0]{\mathbb{1}}

\newcommand{\avg}[1]{\left \langle #1 \right \rangle}
\newcommand{\abs}[1]{\left|#1\right|}
\newcommand{\parS}[1]{\left[#1\right]}
\newcommand{\parO}[1]{\left(#1\right)}
\newcommand{\parC}[1]{\left\{#1\right\}}
\newcommand{\re}[1]{\mathrm{Re}\parC{#1}}
\newcommand{\im}[1]{\mathrm{Im}\parC{#1}}
\newcommand{\mat}[1]{\parO{\begin{matrix}#1\end{matrix}}}
\newcommand{\commutator}[2]{\left[{#1},{#2}\right]}

\newcommand{\ket}[1]{|{#1}\rangle}

\begin{document}

\title{Resonance fluorescence from an artificial atom in squeezed vacuum}
\author{D. M. Toyli}
\thanks{Equal contributors.}
\affiliation{Quantum Nanoelectronics Laboratory, Department of Physics, University of California, Berkeley CA 94720}
\author{A. W. Eddins}
 \thanks{Equal contributors.}
 \affiliation{Quantum Nanoelectronics Laboratory, Department of Physics, University of California, Berkeley CA 94720}
\author{S. Boutin}
\affiliation{D\'epartment de Physique, Universit\'e de Sherbrooke, 2500 boulevard de l'Universit\'e, Sherbrooke, Qu\'ebec J1K 2R1, Canada}
\author{S. Puri}
\affiliation{D\'epartment de Physique, Universit\'e de Sherbrooke, 2500 boulevard de l'Universit\'e, Sherbrooke, Qu\'ebec J1K 2R1, Canada}
\author{D. Hover}
\affiliation{Massachusetts Institute of Technology Lincoln Laboratory, Lexington, MA 02420, USA}
\author{V. Bolkhovsky}
\affiliation{Massachusetts Institute of Technology Lincoln Laboratory, Lexington, MA 02420, USA}
\author{W. D. Oliver}
\affiliation{Massachusetts Institute of Technology Lincoln Laboratory, Lexington, MA 02420, USA}
\affiliation{Department of Physics, Massachusetts Institute of Technology, Cambridge, MA 02139, USA}
\author{A. Blais}
\affiliation{D\'epartment de Physique, Universit\'e de Sherbrooke, 2500 boulevard de l'Universit\'e, Sherbrooke, Qu\'ebec J1K 2R1, Canada}
\affiliation{Canadian Institute for Advanced Research, Toronto, Ontario M5G 1Z8, Canada}
\author{I. Siddiqi}
\affiliation{Quantum Nanoelectronics Laboratory, Department of Physics, University of California, Berkeley CA 94720}
\date{\today}
\begin{abstract}

We present an experimental realization of resonance fluorescence in squeezed vacuum. We strongly couple microwave-frequency squeezed light to a superconducting artificial atom and detect the resulting fluorescence with high resolution enabled by a broadband traveling-wave parametric amplifier. We investigate the fluorescence spectra in the weak and strong driving regimes, observing up to 3.1 dB of reduction of the fluorescence linewidth below the ordinary vacuum level and a dramatic dependence of the Mollow triplet spectrum on the relative phase of the driving and squeezed vacuum fields. Our results are in excellent agreement with predictions for spectra produced by a two-level atom in squeezed vacuum [Phys. Rev. Lett. \textbf{58}, 2539-2542 (1987)], demonstrating that resonance fluorescence offers a resource-efficient means to characterize squeezing in cryogenic environments. 

\end{abstract}
\maketitle
\SIunits[thinspace,thinqspace]

\section{Introduction}

The accurate prediction of the fluorescence spectrum of a single atom under coherent excitation, comprising canonical phenomena such as the Mollow triplet~\cite{Mollow1969, Schuda1974}, is a foundational success of quantum optics. As the Mollow triplet provides a clear signature of coherent light-matter coupling, in recent years such spectra have been widely applied to probe quantum coherence in artificial atoms based on quantum dots~\cite{Xu2007} and superconducting qubits~\cite{Astafiev2010, VanLoo2013}. Resonance fluorescence is predicted to offer analogous spectroscopic means to identify and characterize atomic interactions with squeezed light~\cite{Gardiner1986, Carmichael1987, Carmichael1987a, Parkins1993, Dalton1999}, which is known for its potential to enhance measurement precision in applications ranging from gravitational wave detectors~\cite{Abadie2011} to biological imaging~\cite{Taylor2014}. However, it remains a challenge to demonstrate that exciting atomic systems with squeezed light strongly modifies resonance fluorescence, in part due to the stringent requirement that nearly all the modes in the atom's electromagnetic environment must be squeezed~\cite{Turchette1998}. 

Recently, the circuit quantum electrodynamics architecture has emerged as a compelling platform for studying squeezed light-matter interactions. The low dimensionality of microwave-frequency electrical circuits naturally limits the number of modes involved in atomic interactions~\cite{Wallraff2004}, such that squeezing a single mode can have a significant effect. Experiments have thus demonstrated that microwave-frequency squeezed light can modify the temporal radiative properties of an artificial atom, leading to phase-dependent radiative decay~\cite{Murch2013}. However, these experiments lacked a means to probe the resulting fluorescence, and predictions for the spectrum of resonance fluorescence in both the weak~\cite{Gardiner1986} and strong~\cite{Carmichael1987,Carmichael1987a} driving regimes remain unexplored. Here, we employ two superconducting parametric amplifiers, one to generate squeezing
and one to detect the resulting atomic fluorescence, to perform a systematic study of the dependence of resonance fluorescence on the properties of squeezed vacuum. We observe subnatural fluorescence linewidths and a strong dependence of the Mollow triplet spectrum on the relative phase of the driving and squeezed vacuum fields, in excellent agreement with theoretical predictions. These results enable experimental access to the many theoretical studies on atomic spectra in squeezed reservoirs that have followed~\cite{Parkins1993, Dalton1999}, and demonstrate the utility of resonance fluorescence for the characterization of itinerant squeezed states, an important capability for the development of proposed schemes to enhance qubit readout with microwave-frequency squeezed light~\cite{Barzanjeh2014,Didier2015a}.

\begin{figure}
  \includegraphics[width=\columnwidth]{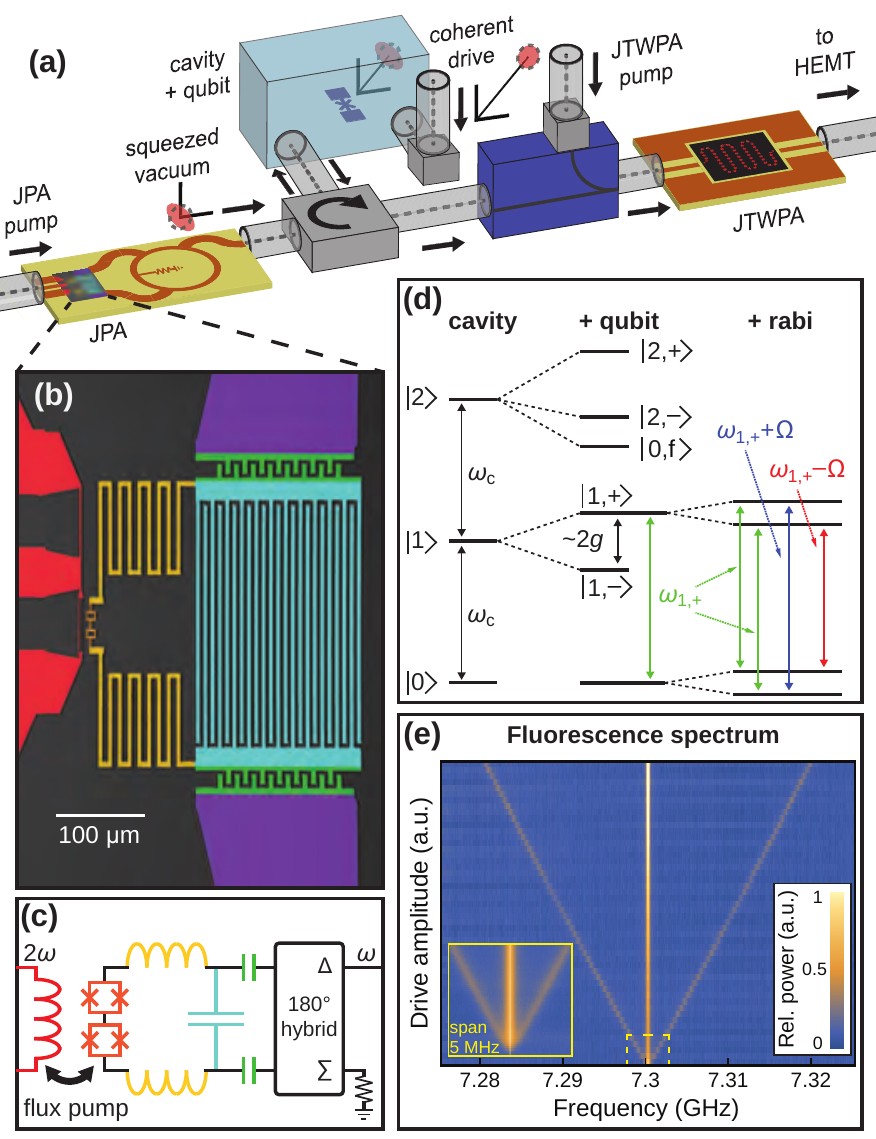}\\
  \caption{\label{intro_fig} (a) Simplified experimental setup. A qubit-cavity system is driven by a coherent drive combined with squeezed vacuum from a Josephson Parametric Amplifier (JPA). The resulting fluorescence is amplified by a Josephson Traveling Wave Parametric Amplifier (JTWPA). (b,c) False-colored photomicrograph and schematic of the JPA. The port (red) at left flux-couples a pump tone to the JPA, producing quadrature squeezing. (d) Coupling to the qubit splits the cavity states into polariton states. Driving Rabi oscillations between $\ket{0}$ and $\ket{1,+}$ splits each state into a doublet. The three distinct transition frequencies produce the Mollow triplet spectrum. (e) Mollow triplet spectra in ordinary vacuum, normalized by the power spectrum with no coherent drive to account for the JTWPA's gain ripple.}
\end{figure}

Our experimental implementation integrates recent advances in parametric amplifiers and superconducting artificial atoms (Fig.~\ref{intro_fig}(a)). We produce itinerant squeezed microwaves using a Josephson parametric amplifier (JPA)~\cite{Castellanos-Beltran2008}, which can be understood as a frequency-tunable LC resonator (Fig.~\ref{intro_fig}(b,c)). A dilution refrigerator cools the experiment to below 30 mK such that the microwave field incident to the JPA is very nearly in its vacuum state. Modulation of the JPA's resonant frequency by a pump tone at twice that frequency squeezes the vacuum fluctuations in one quadrature of the resonator field while amplifying those in the other. Ideally, the JPA's output field is an ellipse in phase space described by the parameters $N$ and $M$, with quadrature phase variances $V(\theta)$ given by
\begin{equation}\label{variances}
V(\theta) = \frac{1}{2}\left[N+M \cos(2(\theta - \varphi))+\frac{1}{2}\right],
\end{equation}
where $M^2 = N(N+1)$. The squeezing axis lies along the angle $\theta = \varphi + \pi/2$ in phase space, at which the minimum quadrature variance occurs (see Appendix~\ref{sec:numerics} for definitions). In practice, $N$ and $M$ are diminished from their ideal values by a factor $\eta < 1$ due to dilution of the itinerant squeezed state with vacuum noise via microwave component loss or other imperfections. 

Our artificial atom consists of a transmon superconducting qubit coupled to an aluminum microwave waveguide cavity~\cite{Paik2011}. As the qubit is nearly resonant with the cavity's fundamental mode, their strong coupling produces well-separated polariton states described by the Jaynes-Cummings Hamiltonian (Fig.~\ref{intro_fig}(d))~\cite{Murch2013,Wallraff2004,Lang2011}. When only two of these states are coupled to squeezed radiation, the system is expected to exhibit the same dynamics as a single two-level atom in squeezed vacuum~\cite{Parkins1993a}. We drive the system via two ports (antennas) coupled to the cavity mode, where one port is coupled more than four times as strongly as the other. Squeezed vacuum from the JPA is directed by a circulator to the cavity's strongly-coupled port, while the weakly-coupled port allows for application of a coherent tone to drive polariton Rabi oscillations. Relatively weak coupling at the latter limits dilution of the squeezing by vacuum fluctuations in the connecting transmission line, such that the two inputs sum to a squeezed vacuum displaced in phase space by the coherent drive. Most of the resulting fluorescence exits the cavity through the strongly coupled port. This fluorescence, in combination with the squeezed radiation reflecting off of the strongly coupled port, then passes through the circulator towards the first stage of amplification.	

To detect the fluorescence with high precision we utilize the recently-developed Josephson traveling wave parametric amplifier (JTWPA)~\cite{Macklin2015}. Previous polariton Mollow triplet studies have relied on cross-correlation of two measurement chains and data processing using a field-programmable gate array (FPGA) to overcome the added noise of following semiconducting amplifiers that would otherwise necessitate prohibitively long averaging times~\cite{Lang2011}. Here, the JTWPA acts as a superconducting preamplifier with near-quantum-limited noise performance that mitigates the effect of this added noise and thus facilitates direct fluorescence measurements with a microwave spectrum analyzer. Moreover, the high dynamic range of the JTWPA ensures that squeezed vacuum power reflecting off of the cavity does not cause compression of the amplified fluorescence spectra. Figure~\ref{intro_fig}(e) plots resonance fluorescence spectra measured in ordinary vacuum over a 50 MHz span as a function of the coherent drive amplitude applied at the upper polariton frequency $(\omega_{1,+})$. For large drive amplitudes, we see a three-Lorentzian Mollow triplet spectrum, reflecting the polariton Rabi oscillations.

\section{Resonance Fluorescence in squeezed vacuum}

We next examine the artificial atom's fluorescence when the polariton is excited only by squeezed vacuum with no coherent drive. In the reflection geometry, non-classical correlations must exist in the driving field to observe any fluorescence peak amidst the broadband noise. To quantitatively analyze the power spectra, we adapt the prediction of Gardiner~\cite{Gardiner1986,Gardiner1987} to the case of a two-port cavity measured in reflection, leading to the expression
\begin{multline}\label{sqspec}
2\pi S_{\mathrm{R}}(\omega) = \frac{ N }{\eta_{\mathrm{c}}} + \gamma \Big[\frac{M-(1- \eta_{\mathrm{c}})N}{2N + 1}\left(\frac{\gamma_y}{\omega^2+\gamma_y^2}\right) \\ -  \frac{M+(1- \eta_{\mathrm{c}})N}{2N + 1}\left(\frac{\gamma_x}{\omega^2 + \gamma_x^2}\right) \Big],
\end{multline}Here $N$ and $M$ describe the squeezed state inside the cavity and $\gamma$ is the radiative linewidth in ordinary vacuum. As predicted for a single-port cavity \cite{Gardiner1986,Gardiner1987}, the spectrum is the sum of a broad background representing reflected noise power, a broad negative Lorentzian with half width $\gamma_x = \gamma(\frac{1}{2} + N + M)$, and a narrow positive Lorentzian with half width $\gamma_y = \gamma(\frac{1}{2} + N - M)$. The negative Lorentzian represents absorption of correlated photon pairs at $\pm \omega$ which are then reemitted at the atomic resonance frequency to produce the positive Lorentzian. Thus if all the incident power is reflected, the areas of the two Lorentzians are equal. However, in practice some power escapes through the second cavity port, which enhances the relative spectral weight of the negative dip and couples in unsqueezed vacuum modes that dilute the squeezing inside the cavity by the factor $\eta_{\mathrm{c}}$. Experimentally, we determine the total linewidth $\gamma = 304$ kHz from fits to the Mollow triplet spectra in ordinary vacuum and the factor $\eta_{\mathrm{c}} = 0.81$ through reflection measurements of the polariton resonance (see Appendix ~\ref{sec::QB-Cav} for details). 

\begin{figure}
  \includegraphics[width=\columnwidth]{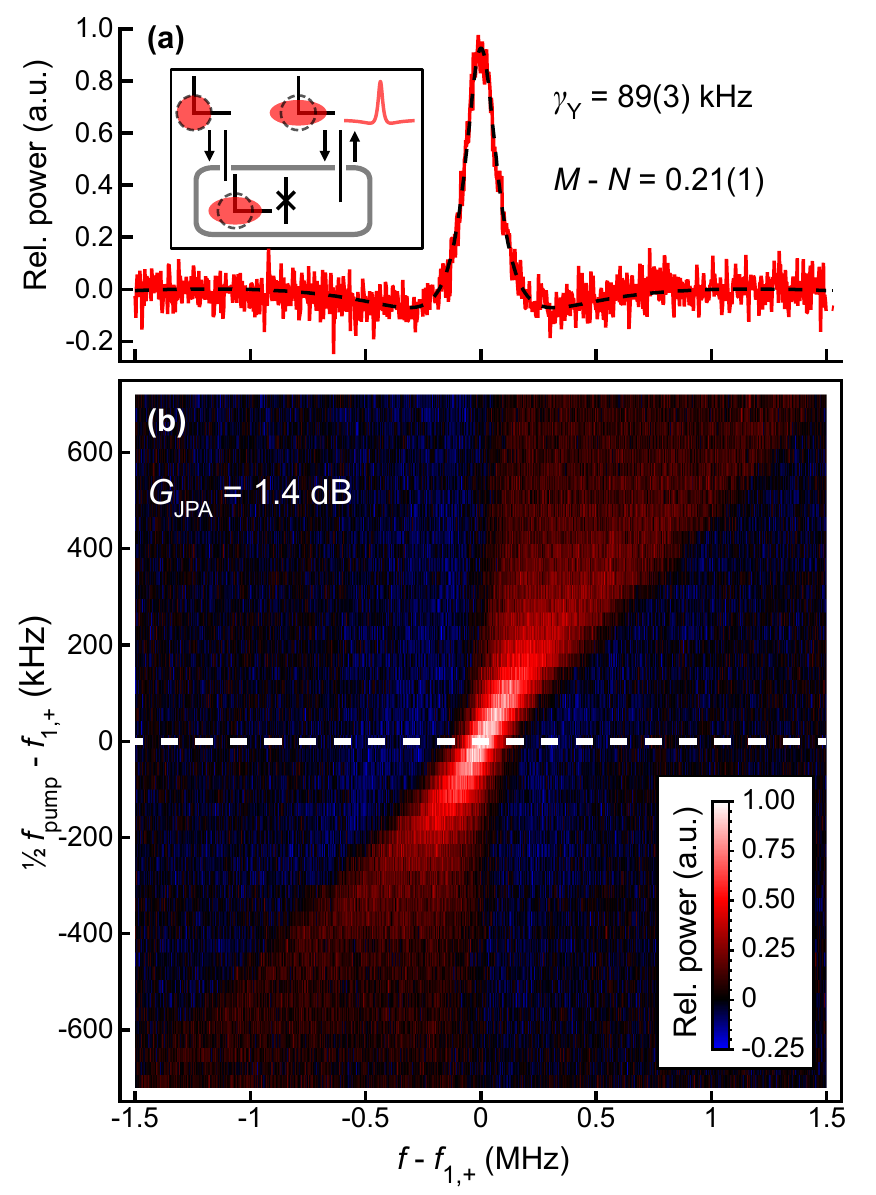}\\
  \caption{\label{sqspec_fig} (a) Reflected power as a function of frequency for excitation with only squeezed vacuum ($G_{\mathrm{JPA}}$ = 1.4 dB). The spectrum is normalized by a measurement in ordinary vacuum to account for the JTWPA's gain ripple. Zero relative power corresponds to the background level of the broadband squeezed power. The red ellipses are phase space representations of the squeezed states inside and outside the cavity as inferred from the fit to Eq. \eqref{sqspec}. (b) Reflected power measured over a constant frequency span as the frequency of the JPA pump is varied. The peak in the spectrum diminishes and broadens as the JPA pump is detuned from the polariton resonance.}
\end{figure}

We observe reflection spectra in good agreement with Eq.~\ref{sqspec}---a representative spectrum for 1.4 dB of JPA signal gain ($G_{\mathrm{JPA}}$, see Appendix ~\ref{sec::JPA}) and a corresponding fit are shown in Fig.~\ref{sqspec_fig}(a). From the fit we determine $\gamma_y$, indicating 2.4 dB of squeezing below the standard vacuum limit for these conditions. While the spectra exhibit negative dips as expected, the small amplitudes and broad spectral range of these features limit the reliable determination of $\gamma_x$ through this measurement. Notably, all of these spectral features rapidly diminish when the squeezer pump is detuned by an amount comparable to $\gamma$ (Fig. \ref{sqspec_fig}(b)). Although much smaller than the squeezer bandwidth, the detuning causes the squeezed state to appear as thermal noise in the frame of the artificial atom. By removing the symmetry of two-mode correlations about the atomic transition, the detuning suppresses the two-photon process by which off-resonant power can be absorbed and then resonantly emitted, thereby suppressing the squeezing induced fluorescence.

\begin{figure}
  \includegraphics[width=\columnwidth]{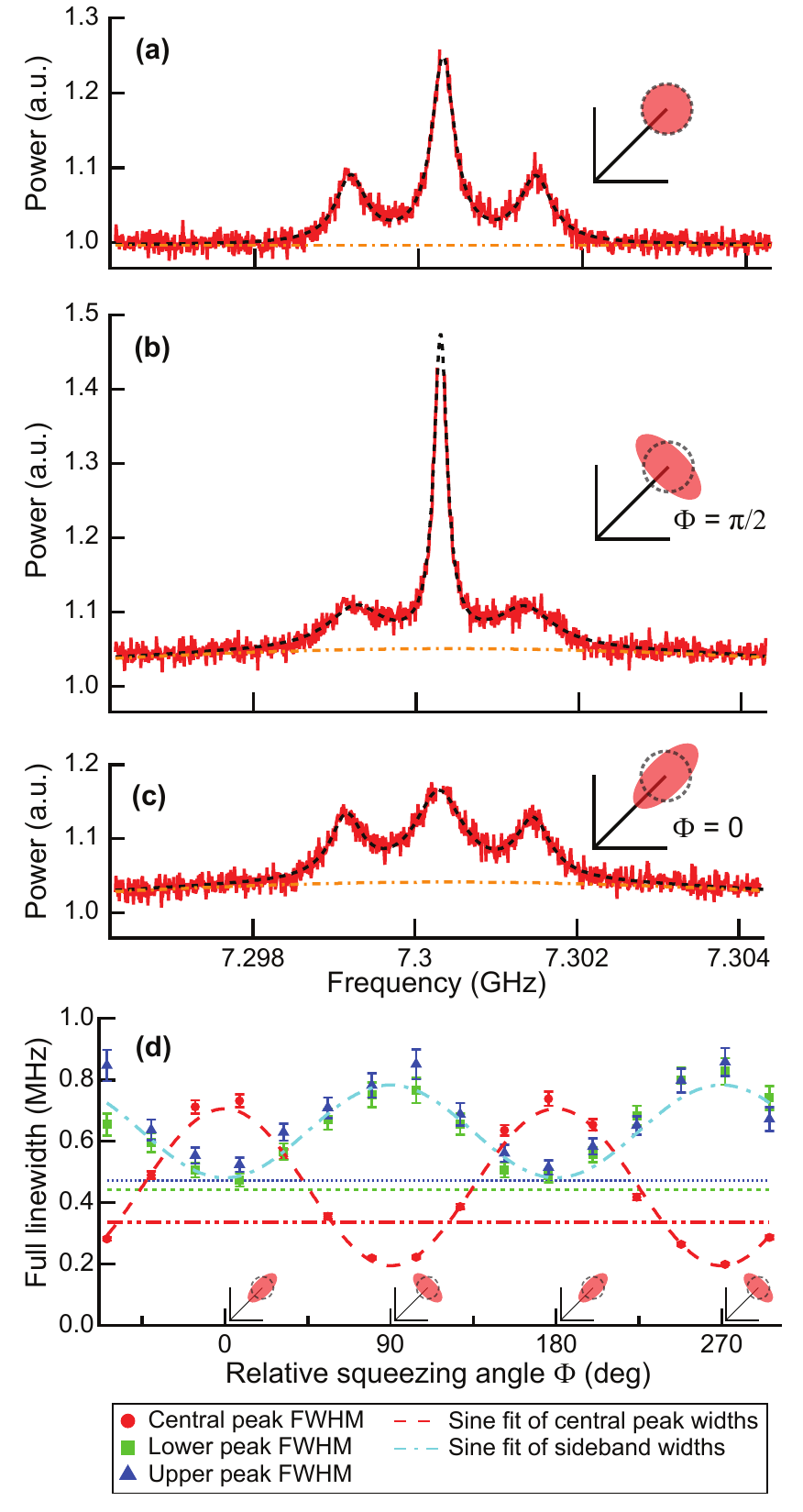}\\
  \caption{\label{Mollow_fig} Reflection spectra for excitation with a coherent drive (a) in ordinary vacuum, (b) with $\Phi = \pi/2$  ($G_{\mathrm{JPA}}$ = 1.5 dB), and (c) with $\Phi = 0$ ($G_{\mathrm{JPA}}$ = 1.5 dB), where $\Phi + \pi/2$ is the phase angle between the coherent drive and the squeezing axis (see Appendix~\ref{sec:numerics} for definitions). The spectra are normalized to account for the JTWPA's gain ripple such that the background level in ordinary vacuum is one; the parabolic backgrounds (orange, dashed) in (b) and (c) result from the reflected JPA noise power. (d) The black dashed curves are fit using an approximate three-Lorentzian model; linewidths of this fit are plotted in d as a function of $\Phi$. Error bars reflect fit uncertainties. The two dashed curves are out-of-phase sine functions fit to the central-peak and sideband linewidths inferred from measurements in squeezed vacuum, while the dashed horizontal lines indicate the corresponding linewidths  inferred from measurements in ordinary vacuum.}
\end{figure}

We next drive the polariton to generate a Mollow triplet spectrum (Fig. \ref{Mollow_fig}(a)). The sideband splitting is kept small so that when the JPA is pumped for gain the squeezing spectrum can be approximated as constant over the Mollow triplet spectral range. As the polariton's coherent drive and the JPA's flux pump are generated from the same microwave source, their relative phase can be controllably varied (Fig.~\ref{Mollow_fig} insets). This configuration facilitates observation of the Mollow triplet's strong dependence on this relative phase as predicted by Carmichael, Lane, and Walls~\cite{Carmichael1987,Carmichael1987a}: at modest JPA gains all three peaks can be resolved with the center peak oscillating between subnatural (Fig. \ref{Mollow_fig}(b)) and supernatural (Fig.~\ref{Mollow_fig}(c)) half linewidths of $\gamma_y$ and $\gamma_x$, respectively. To further demonstrate the phase dependence, we first fit the data with an approximate model of three Lorentzian peaks; example fits appear as black dashed curves in Fig. \ref{Mollow_fig}(a-c). The results of these fits exhibit the expected out-of-phase oscillations of the center peak and sideband linewidths as a function of this relative phase (Fig.~\ref{Mollow_fig}(d)).  

\section{Squeezing characterization}

While the approximate Lorentzian model clearly illustrates the phase dependence of the spectra, we quantitatively characterize the squeezing levels through fits to analytic expressions for the power spectra having both Lorentzian and small non-Lorentzian dispersive constituents (Appendix~\ref{sec:numerics}, Eq.~(B20)). These expressions extend the results of Refs.~\cite{Carmichael1987,Carmichael1987a} to arbitrary relative phases and do not assume observation through unsqueezed vacuum modes. We find that these analytical expressions are able to accurately model the measured power spectra even at large $G_{\mathrm{JPA}}$ when the sidebands are significantly broadened and the frequency-dependence of the squeezer background becomes prominent.

\begin{figure}
  \includegraphics[width=\columnwidth]{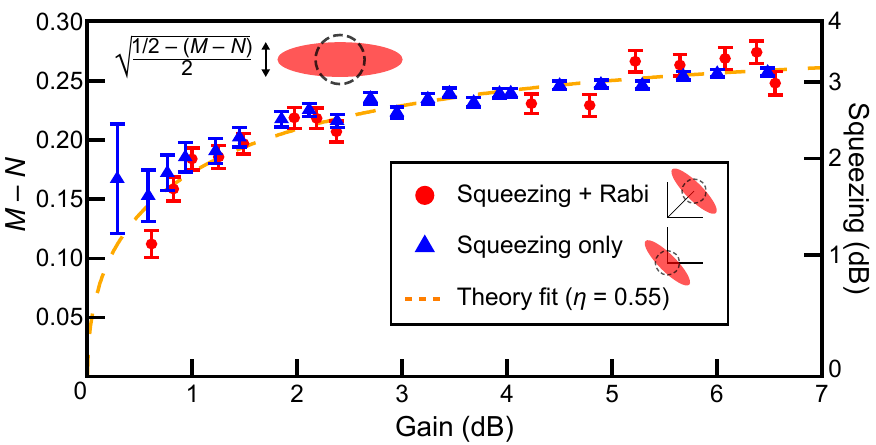}\\
  \caption{\label{sq_fig} The squeezing level in the artificial atom's environment as a function of gain $G_{\mathrm{JPA}}$. The quantity $M-N$ determines the minimal quadrature variance of a squeezed state as indicated in the inset. The values of $M-N$ are inferred from measurements of the fluorescence spectrum in squeezed vacuum with (red) and without (blue) a coherent Rabi drive. The error bars reflect fit uncertainties. The measurements with no Rabi drive give a more reliable indication of the squeezing level as they are less sensitive to background effects and experimental drift.  The dashed orange curve is a one-parameter fit to the blue data points yielding $\eta = 0.55$.}
\end{figure}

Figure~\ref{sq_fig} plots $M - N$, which indicates the squeezing level, as a function of $G_{\mathrm{JPA}}$, comparing results from fits of the Mollow triplet spectra (red points) to corresponding results from fits of spectra measured with no Rabi drive (blue points). Both measurements of $M - N$ are well-described by a one-parameter fit with overall efficiency $\eta = 0.55$ (orange dashed line), comprising losses due to both $\eta_{\mathrm{c}}$ and component loss (see Appendix \ref{sec::QB-Cav}). For the range of $G_{\mathrm{JPA}}$ measured here, we observe cavity squeezing levels as high as 3.1 dB below the standard vacuum limit; factoring out cavity losses corresponding to $\eta_{\mathrm{c}}$ suggests up to 4.4 dB of squeezing in the itinerant squeezed state incident to the cavity. While the results of the measurements with and without a Rabi drive are in reasonable agreement, the simplicity, speed, and relative insensitivity to background features of the measurements with no Rabi drive recommend that technique as a resource-efficient detector of squeezing, which could be implemented in a single-port cavity to further improve precision.

Both generating and characterizing the squeezing in a superconducting qubit's environment are of central importance to proposed schemes for enhancing qubit measurement via interferometric readout~\cite{Barzanjeh2014,Didier2015}. The ratiometric methods presented here achieve this characterization through spectral measurements that are considerably simpler to implement than alternative detection methods that require time-domain qubit control~\cite{Murch2013}. Moreover, as subnatural fluorescence linewidths are also expected to occur in two-mode squeezed vacuum~\cite{Parkins1990a}, these metrological techniques are similarly relevant to measurement schemes employing two-mode squeezed light~\cite{Didier2015a}.

More generally, our results validate the canonical predictions for resonance fluorescence in squeezed vacuum and exemplify the JTWPA's broad utility in microwave quantum optics. These experimental techniques directly enable investigations of numerous predicted phenomena including fluorescence in non-Markovian squeezed reservoirs~\cite{Parkins1990} and in parameter regimes expected to produce qualitatively distinct spectra~\cite{Swain1994}. Finally, while this area of experimental study has long been led by theoretical prediction, combining the fluorescence detection techniques demonstrated here with the wide range of atomic environments realizable in circuit quantum electrodynamics enables the study of squeezing's impact on resonance fluorescence in previously unforeseen settings such as multimode strong coupling \cite{Sundaresan2015} and the ultrastrong coupling regime~\cite{Niemczyk2010}.

\begin{acknowledgments}
The authors thank E. Flurin, C. Macklin, and L. Martin for useful discussions. Work was supported in part by the Army Research Office (W911NF-14-1-0078); the Office of Naval Research (N00014-13-1-0150); NSERC; the Office of the Director of National Intelligence (ODNI), Intelligence Advanced Research Projects Activity (IARPA), via MIT Lincoln Laboratory under Air Force Contract FA8721-05-C-0002; and a Multidisciplinary University Research Initiative from the Air Force Office of Scientific Research MURI grant no. FA9550-12-1-0488. The views and conclusions contained herein are those of the authors and should not be interpreted as necessarily representing the official policies or endorsements, either expressed or implied, of ODNI, IARPA, or the U.S. government. The U.S. government is authorized to reproduce and distribute reprints for governmental purpose notwithstanding any copyright annotation thereon. A.W.E. acknowledges support from the Department of Defense through the NDSEG fellowship program.
\end{acknowledgments}

\appendix

\section{Experimental Methods}

\subsection{Qubit-cavity implementation}\label{sec::QB-Cav}

The qubit was composed of an aluminum transmon circuit fabricated on double side polished silicon characterized by $E_{\mathrm{C}}/h$ = 200 MHz and $E_{\mathrm{J}}/h$ = 33.1 GHz. The transmon circuit was coupled to a 3D aluminum cavity with resonance frequency $\omega_{\mathrm{c}}/2\pi$ = 7.1051 GHz at rate $g/2\pi$ = 202 MHz. We calculate the frequency of the bare qubit to be 7.091 GHz. To bring the cavity close to resonance with the qubit, we placed extra silicon chips in the cavity as a low-loss dielectric based on finite-element simulation. Most of the final detuning of 14 MHz was due to aging of the qubit between experimental cycles. 

As mentioned in the main text, coupling to unsqueezed vacuum modes dilutes the squeezing in the artificial atom's environment from that which is produced at the JPA output. We probe the cavity contribution to this dilution by measuring the $|1,+\rangle$ polariton resonance in reflection through the strongly coupled cavity port. Through this port, we measure external and internal quality factors of $Q_{\mathrm{1+,ext}} = 26,500$ and $Q_{\mathrm{1+,int}} = 110,000$, the latter characterizing the total loss rate due to the extra silicon substrates and the weak coupling of the cavity's second port. This coupling to unsqueezed modes limits the achievable squeezing, corresponding to $\eta_{\mathrm{c}} = Q_{\mathrm{1+,int}}/(Q_{\mathrm{1+,ext}} + Q_{\mathrm{1+,int}})  = 0.81$. Given that we experimentally observe $\eta = \eta_{\mathrm{loss}}\eta_{\mathrm{c}} = 0.55$, we infer $\eta_{\mathrm{loss}} = 0.68$. This number is in reasonable agreement with expectations for the combined effect of internal paramp loss and component loss in the circulator, hybrid, and cables. In future experiments, $\eta_{\mathrm{c}}$ could be optimized by reducing the coupling of the weakly coupled port, with $\eta_{\mathrm{loss}}$ limiting the minimum achievable linewidth.

Approximating our multilevel qubit-cavity system as a two-level state requires the use of squeezed light with sufficiently narrow bandwidth to avoid exciting the system's higher levels. For a resonant Jaynes-Cummings Hamiltonian this imposes the restriction that $\kappa_{\mathrm{JPA}} < 0.6g$~\cite{Parkins1993a}. However, it is important to account for the fact that the higher levels of the transmon qubit can perturb the frequencies of the dressed states. We directly characterized the relevant transition frequencies using two-tone spectroscopy~\cite{Fink2008}. When resonantly driving the $|0,g\rangle$ to $|1,+\rangle$ transition with a coherent tone we find that the closest transition is the $|1,+\rangle$ to $|2,+\rangle$ transition at 7.262 GHz, corresponding to a detuning of -38 MHz. The JPA was designed to have a single-side bandwidth smaller than this detuning.

To confirm that our qubit-cavity system was well-thermalized to the 30 mK base stage of our dilution refrigerator and that thermal photons were not contaminating the vacuum environment, we measured the excited-state population of the polariton states using a Rabi population measurement \cite{Geerlings2013,Jin_PRL_2015}.  These measurements indicate the excited state population was less than $1.2\%$, corresponding to $N_{\mathrm{th}} < 0.01$, thus negligibly impacting the squeezing levels quoted in the main text. 

\subsection{Josephson parametric amplifier}\label{sec::JPA}

The JPA used in this work was designed to provide a high degree of squeezing despite operating over a narrow bandwidth. The JPA's squeezing performance is expected to inversely correlate with the dimensionless nonlinearity $\Lambda/\kappa_{\mathrm{JPA}}$~\cite{Eichler2014}, where $\Lambda$ is the Kerr nonlinearity and $\kappa_{\mathrm{JPA}}$ is the resonator bandwidth. Here $\kappa_{\mathrm{JPA}}$ was intentionally kept small ($\kappa_{\mathrm{JPA,external}}/2\pi$  = 21 MHz when $\omega_{\mathrm{JPA}} = \omega_{1,+}$) through the use of input coupling capacitors to avoid exciting higher transitions in the qubit-cavity system. To produce a weak nonlinearity, $\Lambda$ was reduced through the incorporation of geometric inductance, which reduces the Josephson junction participation ratio ($p = L_\mathrm{J}/L_{\mathrm{total}}$), and through the use of a two SQUID array. The former reduces $\Lambda$ by $p^3$, whereas the latter reduces $\Lambda$ by $1/N_{\mathrm{SQUID}}^2$ for fixed $L_\mathrm{J}$, where $N_{\mathrm{SQUID}}$ is the number of SQUIDs. With these modifications, we estimate $\Lambda/\kappa_{\mathrm{JPA}} \sim 5 \times 10^{-5}$ at the polariton frequency. For comparison, $\Lambda/\kappa_{\mathrm{JPA}} \sim 6 \times 10^{-3}$ for an ideal lumped-element JPA consisting of a capacitance shunted by a single SQUID with a 100 Ohm input impedance (set by the $180^{\circ}$ hybrid launch). In addition, the amplifier was pumped for gain by modulating the flux through the two SQUID loops at twice the resonator frequency. Relative to alternative methods such as double-pumping~\cite{Murch2013}, flux-pumping allowed us to minimize the number of passive components between the JPA and the artificial atom. All of the JPA circuitry, including the flux pump input and the hybrid, was shielded by an aluminum box and cryoperm at the base stage of the dilution refrigerator.

For each JPA pump amplitude, the JPA power gain ($G_{\mathrm{JPA}}$) was characterized using a vector network analyzer. These $G_{\mathrm{JPA}}$ values refer to phase preserving gain of the amplifier, as the gain was characterized for a signal slightly detuned such that the measurement bandwidth was small compared to the detuning. For an ideal squeezed state, $G_{\mathrm{JPA}}$ is related to the power gain in the amplified and squeezed quadratures by the expression
\begin{equation}\label{gain}
2\left(N\pm M+\frac{1}{2}\right) = \left(\sqrt{G_{\mathrm{JPA}}} \pm \sqrt{G_{\mathrm{JPA}}-1} \right)^2.
\end{equation}

\subsection{Data acquisition and analysis}\label{sec::DataAc}

\begin{figure*}
  \begin{center}
\includegraphics[angle = 0, width = .70\textwidth]{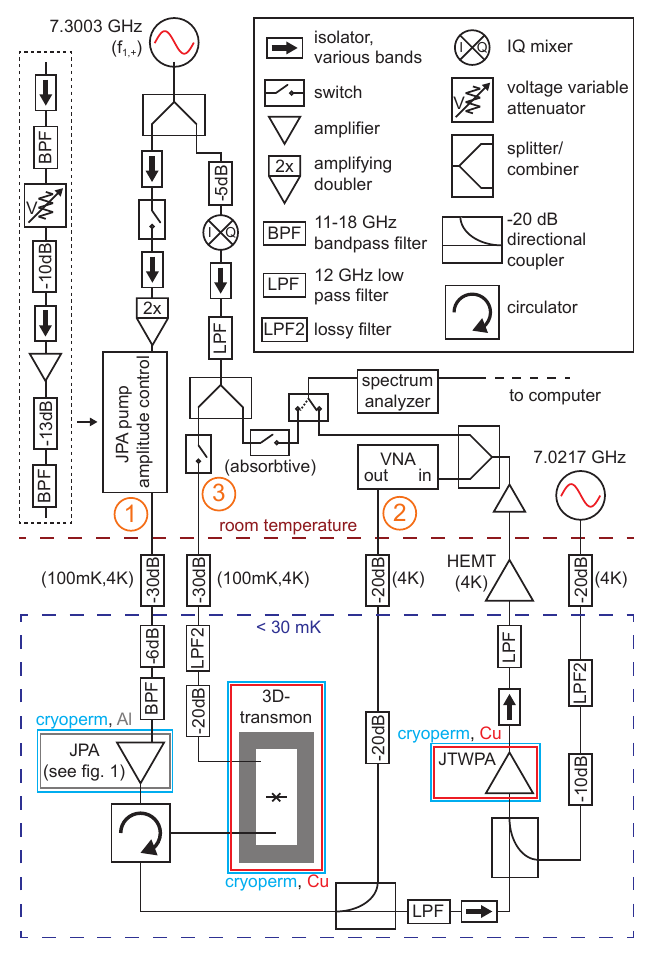}
\end{center}
\caption{\label{fig:setup} Experimental setup diagram. The JPA pump tone, which enters the cryostat at (1), is generated through a frequency doubler followed by a voltage variable attenuator that allows programmatic control of the JPA gain. The gain can be measured by the VNA at (2). Power from the same generator is split off before doubling to create the phase-locked coherent drive at (3). The relative phase of this drive is controlled through DC voltages on the IQ mixer. Each time the phase is stepped, the two switches just before the spectrum analyzer are thrown such that the analyzer samples the drive power, which is fed back on to ensure power flatness versus phase. These switches are toggled back before measurement of fluorescence spectra on the spectrum analyzer.}  
\end{figure*}

Fig.~\ref{fig:setup} displays a full schematic of the experimental setup. The polariton drive and the JPA flux pump were phase locked by generating them from the same microwave source and doubling the flux pump frequency. This allowed the relative phase between the two tones to be controllably varied by adjusting the bias voltages on an IQ mixer. All spectra were directly acquired using a microwave spectrum analyzer. As our analysis focuses on the incoherent portion of the fluorescence spectrum, we kept the spectrum analyzer resolution bandwidth much smaller than any of the relevant spectral linewidths such that the coherent portion of the spectrum could be omitted by dropping a small number of points around the polariton frequency. In addition, each spectrum was normalized by an equivalent measurement with both the JPA pump and coherent drive off in order to eliminate background trends originating from the gain ripple of the JTWPA~\cite{Macklin2015} and following amplifiers. We note that, as we are working in the photon blockade regime, we expect the fluorescence intensity in the Mollow triplet measurements (not including the background squeezed noise power) to be approximately $\hbar\omega\gamma = -140$ dBm. For all measurements the JTWPA was pumped to provide roughly 20 dB of power gain.

After normalizing the data, we determined the squeezing levels using the analytic fit expressions outlined in the main text. Here we describe additional details of the fitting procedures. For \mbox{Fig.~\ref{sqspec}(a)}, the squeezing level was inferred by fitting the data to Eq. \ref{sqspec} from the main text including an overall scaling and offset. In addition, the fit includes a parabolic background term to account for the frequency-dependence of the squeezed noise power produced by the JPA. Although small in \mbox{Fig. \ref{sqspec}(a)}, this background becomes important when analyzing such spectra at large gains as is done to determine the squeezing levels presented in \mbox{Fig.~\ref{sq_fig}}. To determine the total linewidth ($\gamma$) used to constrain the inputs to Eq. \ref{sqspec} in the main text, we fit the Mollow triplet spectra in ordinary vacuum using the methodology described in Appendix~\ref{sec:numerics} to account for dispersive spectral features and the weakly coupled cavity port. In \mbox{Fig.~\ref{Mollow_fig}}, to illustrate the phase dependence of the Mollow triplet, the spectra were approximated as the sum of three Lorentzians and a parabolic background. Approximating the peaks neglects the aforementioned dispersive features. Therefore, to more rigorously characterize the squeezing level through this measurement, we perform nonlinear regression using the analytic expressions in Appendix~\ref{sec:numerics}. In \mbox{Fig.~\ref{sq_fig}} we focus on characterizing $M-N$ by measuring spectra at four relative phases between the coherent drive and squeezed vacuum fields near the phase that minimizes the center peak linewidth. For each gain point, the four spectra were fit with joint parameters; only background and scaling terms were allowed to vary among the four traces to compensate for small experimental drift. As in Fig.~\ref{sqspec}, for these fits $\gamma$ was independently determined by fitting the Mollow triplet spectrum in ordinary vacuum.

\section{Numerical modeling of fluorescence spectra}\label{sec:numerics}
\begin{figure*}
\includegraphics[width = \textwidth]{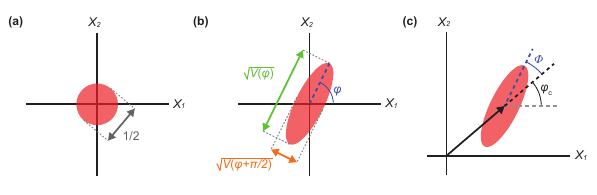}
\caption{\label{fig:defs} States of light can be fully described by quasiprobability distributions (here Wigner functions) in phase space, where $X_1$ and $X_2$ are the two quadratures of the light field. \textbf{a}, The vacuum state is described by a circular Gaussian distribution of width $1/2$ centered at the origin. The red disk indicates the area in which the distribution is greater than its half-maximum value. \textbf{b}, The distribution for a squeezed vacuum state has an increased width along the amplification axis at angle $\varphi$ and a decreased width along the squeezing axis at angle $\varphi + \pi/2$. A slice of this distribution taken through the origin at angle $\theta$ has a variance given by Eq. 1 of the main text, with the amplified variance given by $1/2(N + M + 1/2)$ and the squeezed variance given by $1/2(N - M + 1/2)$. Note an angular factor of 2 has been introduced in Eq.~\ref{variances} in the main text compared to Refs. ~\cite{Carmichael1987,Carmichael1987a} to simplify the phase space picture. \textbf{c}, Similarly, $\Phi$, as used in Fig.~\ref{Mollow_fig} of the main text, indicates the angle of the amplification axis of a displaced squeezed state relative to the displacement angle, $\varphi_c$. Experimentally, $\varphi_c$ is set by the phase of the coherent Rabi drive.}  
\end{figure*}

In the papers of Carmichael, Lane and Walls~\cite{Carmichael1987,Carmichael1987a}, an expression for the resonance fluorescence spectrum of a two-level atom driven by broadband squeezed light is given for the limiting cases $2\Phi=0, \pi$, where $\Phi$ is the relative phase between the Rabi drive and the squeezing angle. Note that here, as depicted in Fig. \ref{fig:defs}, an angular factor of two has been introduced compared to Refs.~\cite{Carmichael1987,Carmichael1987a} to simplify the phase space picture. In order to analyze the experimental data for an arbitrary phase $\Phi$, we generalize these analytical expressions.
In Sec.~\ref{sec::blochEqs}, we define the notation used and give the optical Bloch equations.
 In Sec.~\ref{sec::spectrum}, we obtain the resonance fluorescence spectrum using the Bloch equations and the quantum regression theorem. Finally, in Sec.~\ref{sec::reflection}, we discuss the calculation of the reflection spectrum and the broadband squeezing approximation.

\subsection{Definitions and steady-state solutions of the optical Bloch equations}\label{sec::blochEqs}
When an atom is strongly coupled to a cavity with dipole coupling $g$, the eigenstates of the system are the dressed-states called polaritons formed by a superposition of the bare atom, cavity states: $\ket{+,n}=[\ket{g,n}+\ket{e,n-1}]/\sqrt{2},\ket{-,n}=[\ket{g,n}-\ket{e,n-1}]/\sqrt{2}$, with vacuum Rabi splitting $2g\sqrt{n}$. It has been shown that for a strong $g$ each vacuum-Rabi resonance behaves as a two-level system~\cite{Tian1992}. As described in the main text, the  system is driven at the upper polariton $\ket{\mathrm{+,1}}=[\ket{g,1}+\ket{e,0}]/\sqrt{2}$ frequency.  As long as no other dressed-states are excited, the ground state  $\ket{g,0}$ and $\ket{\mathrm{+,1}}$ behave as an effective two-level system. The next closest transition frequency corresponds to the $\ket{+,1}$ to $\ket{+,2}$ transition, with $\omega_{+,2}-\omega_{+,1}=(2-\sqrt{2})g\sim 0.6g$. As a result, the two-level approximation holds for $\kappa_{\mathrm{JPA}},\Omega\ll 0.6g$, where $\kappa_{\mathrm{JPA}}$ is the linewidth of the JPA acting as the source of squeezing  and $\Omega$ the coherent drive amplitude. 

Working in this limit, we only consider the two-level system $\{\ket{g,0},\ket{+,1}\}$, driven in resonance in a broadband squeezed vacuum environment, $\kappa_{\mathrm{JPA}}\gg\Omega,\gamma$ with $\gamma$ the linewidth of the two-level system. With this approximation and working in the rotating frame of the drive, the master equation describing the two-port system is~\cite{Gardiner2004}
\begin{align}
\begin{split}
	\dot{\rho}&=
	\frac{i\Omega}{2}[\sigma_+ +\sigma_-,\rho]
	+\gamma_{\mathrm{int}}\mathcal{D}[\sigma_-]\rho
	\\ &
	+\gamma_{\mathrm{ext}} (\tilde{N}+1)\mathcal{D}[\sigma_-]\rho
	+\gamma_{\mathrm{ext}}  \tilde{N} \mathcal{D}[\sigma_+]\rho
	\\ &
	-\gamma_{\mathrm{ext}} \tilde{M}\mathcal{S}[\sigma_+]\rho
	-\gamma_{\mathrm{ext}}  \tilde{M}^*\mathcal{S}[\sigma_-]\rho,
	\label{eq::ME}
\end{split}
\end{align}
%
where $\tilde{N}$ and $\tilde{M}$ are the second-order moments of the incoming squeezed field, 
$\mathcal{D}[A]\rho = A\rho A^\dag - \frac{ 1 }{2} (A^\dag A \rho + \rho A^\dag A)$ and
$\mathcal{S}[A]\rho = A\rho A - \frac{ 1 }{2} (A^2 \rho + \rho A^2)$ 
are superoperators.

%
%
In order to simplify the notation, we use the quantum efficiency $\eta_c = \gamma_{\mathrm{ext}}/(\gamma_{\mathrm{int}}+\gamma_{\mathrm{ext}})$ to rewrite the master equation as a two-level system coupled to a single port
\begin{equation}
	\begin{split}
	\dot{\rho}
	&=
	\frac{i\Omega}{2}[\sigma_++\sigma_-,\rho]+\gamma( N+1)\mathcal{D}[\sigma_-]\rho
	\\ &\quad
	+\gamma  N \mathcal{D}[\sigma_+]\rho
	-\gamma M\mathcal{S}[\sigma_+]\rho
	-\gamma  M^*\mathcal{S}[\sigma_-]\rho,
	\label{eq::ME2}
\end{split}
\end{equation}
with effective parameters $N = \eta_c \tilde{N}$ and $M  = \eta_c \tilde{M}$ such that $\gamma_{\mathrm{ext}} \tilde{N} = \gamma N$.

The optical Bloch equations in the presence of squeezing for the effective single port system can be obtained from Eq.~\eqref{eq::ME2}, leading to
\begin{equation}
\begin{split}
\langle\dot{\sigma}_x\rangle&=-\gamma_+\langle\sx\rangle+\gamma_M\langle\sy\rangle\\
\langle\dot{\sigma}_y\rangle&=\gamma_M\langle\sx\rangle-\gamma_-\langle\sy\rangle-\Omega\langle\sz\rangle\\
\langle\dot{\sigma}_z\rangle&=\Omega\langle\sy\rangle-\gamma_N\langle\sz\rangle-\gamma,
\label{eq::blochEqs}
\end{split}
\end{equation}
where $\sigma_{x,y,z}$ are the standard Pauli matrices, $\gamma_{\pm} = \gamma\parO{N \pm \abs{M}\cos 2\Phi +\frac{1}{2} }$, $\gM = \gamma \abs{M} \sin 2\Phi$ and $\gN = \gamma \parO{2N+1}$.
The steady-state solutions of these equations are
\begin{equation}
	\begin{split}
		 \avg{\sx}_{ss} = \frac{\Omega \gamma \gM}{\gN \gNM^2+ \Omega^2 \gp},
		  \quad
		  \\
		 \avg{\sy}_{ss} = \frac{\Omega \gamma \gp}{\gN \gNM^2 + \Omega^2 \gp},
		 \quad
		  \\
		 \avg{\sz}_{ss} = -\frac{\gamma \gNM^2}{\gN \gNM^2 + \Omega^2 \gp},
	\end{split}
\end{equation}
with the rate $\gNM^{^2} = \gamma^{^2} \parS{ \parO{N+\frac{1}{2}}^2 - M^2}$.
\subsection{Resonance fluorescence of a driven two-level system in a squeezed environment}\label{sec::spectrum}
The resonance fluorescence spectrum is obtained from the Fourier transform of a time-domain correlation function~\cite{Gardiner2004}
\begin{equation}
	S(\omega) =  \frac{1}{\pi}\re{
		\int_{0}^{\infty} \drm t\,  \avg{ \sp(t) \sm(0)}_{ss} \erm^{i \omega t}
	}.\label{eq::spectrum}
\end{equation}
Noting that the correlation function of interest can be expressed as~\cite{Carmichael1987a}
\begin{multline}
	\avg{ \sp(t) \sm(0)}_{ss}   =	[
		\avg{\sx(t)\sx(0)}_{ss}
		+
		\avg{\sy(t)\sy(0)}_{ss}
		\\ +i \avg{\sy(t)\sx(0)}_{ss}
		-i\avg{\sx(t)\sy(0)}_{ss}
	],
	\label{eq::corrFunction}
\end{multline}
we use Eq.~\eqref{eq::blochEqs} and the quantum regression formula~\cite{Gardiner2004} to obtain the linear system of equations,
\begin{multline}
	\partial_{t} \mat{
		\avg{\sx(t)\sxy(0)}_{ss}
		\\
		\avg{\sy(t)\sxy(0)}_{ss}
		\\
		\avg{\sz(t)\sxy(0)}_{ss}
	}
	=
	\\ B \mat{
		\avg{\sx(t)\sxy(0)}_{ss}
		\\
		\avg{\sy(t)\sxy(0)}_{ss}
		\\
		\avg{\sz(t)\sxy(0)}_{ss}
	}
	- \gamma
	\mat{0\\0\\ \avg{\sxy}_{ss} },
	\label{eq:eqSystem}
\end{multline}
with the matrix
\begin{equation}
	B = \mat{
		-\gp & \gM & 0 \\
		\gM & -\gm & -\Omega \\
		0 & \Omega & -\gN  
	}.
\end{equation}
The results of Ref.~\cite{Carmichael1987,Carmichael1987a} are obtained for $\gM =0$. Here, we generalize these results.

In order to solve this set of linear equations for arbitrary time $t$, we make use of the Laplace transform:  $\Lm \parC{f(t)} = \int_{0}^{\infty}  \erm^{-s t} f(t) \, \drm t$. Performing this transformation on Eq.~\eqref{eq:eqSystem} we obtain
%
\begin{multline}
	\Lm\parC{\mat{
		\avg{\sx(t)\sxy(0)}_{ss}
		\\
		\avg{\sy(t)\sxy(0)}_{ss}
		\\
		\avg{\sz(t)\sxy(0)}_{ss}
	}}
	=
	\\ \parO{s\id -B}^{-1}
		\mat{
			\avg{\sx\sxy}_{ss}
			\\
			\avg{\sy\sxy}_{ss}
			\\
			\avg{\sz\sxy}_{ss}- \frac{\gamma}{s}\avg{\sxy}_{ss}
		}
		%
	,
\end{multline}
with the relations
\begin{widetext}
\begin{equation}
	\parO{s \id - B}^{-1} = \frac{1}{D(s)} \mat{
		\parO{s+\gm}\parO{s+\gN} + \Omega^2 & \gM \parO{s+\gN} & - \gM \Omega \\
		\gM\parO{s + \gN} & \parO{s+\gN}\parO{s+\gp} & - \parO{s+\gp}\Omega \\
		 \gM \Omega & \parO{s +\gp} \Omega & \parO{s+\gm}\parO{s+\gp}-\gM^2 
	},
\end{equation}
\begin{equation}
	D(s)
	=
	\parO{s+ \gN} \parS{
	 \parO{s+\gm} \parO{s+\gp}-\gM^2 
	}
	+ \parO{s+\gp} \Omega^2.
	\label{eq:cubic} 
\end{equation}
\end{widetext}

\begin{figure*}
\includegraphics[width = \textwidth]{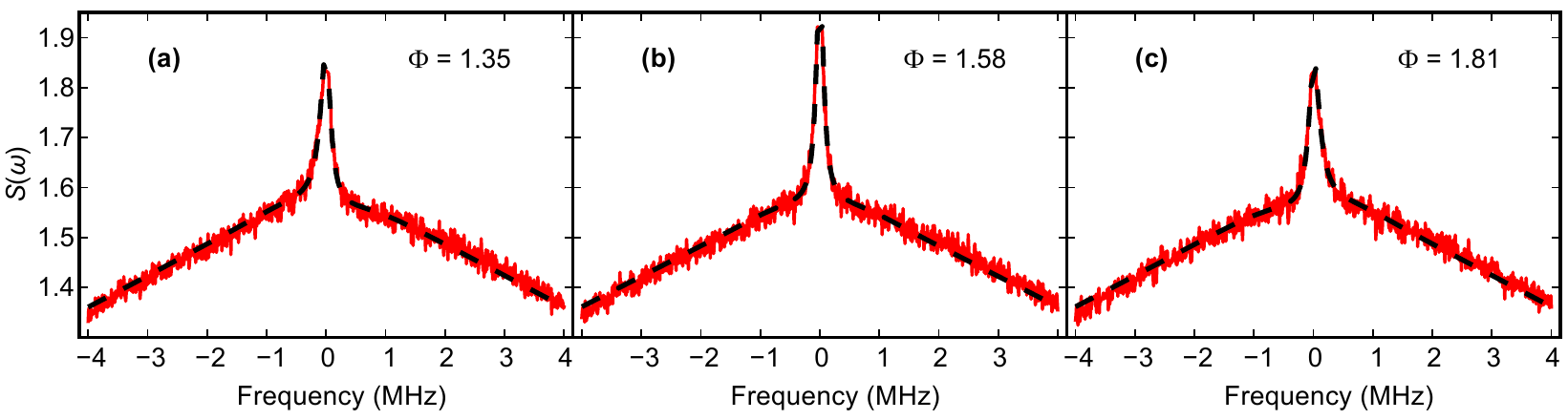}
\caption{\label{fig:spectra} Mollow triplet spectra measured at $G_{\mathrm{JPA}}$ = 6.6 dB. The spectra are measured near $\Phi = \pi/2$, the relative phase between the squeezed vacuum and coherent drive fields that minimizes the center peak linewidth. Even though the Rabi frequency (1.2 MHz as in Fig. 3) is small compared to the measured frequency range, the sidebands are broadened such that they are no longer apparent. The dashed black line represents the best fit to the power spectra, taking into account both the fluorescence spectrum and the frequency-dependent squeezer background. The spectra are simultaneously fit with $\gamma$ fixed by measurements of the Mollow triplet in ordinary vacuum and also constrained by the relative phase difference between each measurement. From these measurements, we infer $M - N$ = 0.24 for these conditions.}  
\end{figure*}

In order to obtain the time-domain solution, an inverse Laplace transform must be applied, which requires the roots $\lambda_{0,1,2}$ of the cubic polynomial $D(s)$.
In the limiting cases $2\Phi=\{0,\pi\}$, the rate $\gamma_M$ is zero and the roots of $D(s)$ are easily obtained as~\cite{Carmichael1987,Carmichael1987a}
\begin{equation}
\begin{split}
	\lambda_0 &= - \gp,
	\\
	\lambda_{1,2} &= -\frac{\gm+\gN}{2} \pm \frac{1}{2}\sqrt{\parO{\gm-\gN}^2 - 4 \Omega^2}.
\end{split}
\end{equation}
However, for an arbitary phase $\Phi$, finding analytical solutions requires the cubic roots formula. As these analytical expressions are rather cumbersome, but easily found, they are not reproduced here. With these expressions, $D(s)$ can be expressed as
\begin{equation}
	D(s) = \parO{s - \lr} \parO{s-\lc}\parO{s-\ld}.
\end{equation}
Using this factorization, the inverse Laplace transform reads
\begin{multline}
	{ f_n(t)} = \Lm^{-1}\parC{\frac{s^n}{D(s)}} = \\
	C_0 \lr^n \erm^{\lr t} + C_1 \lc^n \erm^{\lc t} + C_2 \ld^n \erm^{\ld t} - \frac{\delta_{n,-1}}{\lr\lc\ld}
\end{multline}
with $n \in \parC{-1,0,1,2}$ and where we have defined the coefficients
\begin{equation}
\begin{split}
	C_0 = \frac{1}{\parO{\lr-\lc}\parO{\lr-\ld}},
	\\
	C_1 = \frac{1}{\parO{\lc-\lr}\parO{\lc-\ld}},
	\\
	C_2 = \frac{1}{\parO{\ld-\lr}\parO{\ld-\lc}}.
\end{split}
\end{equation}
Hence time-domain solutions are 
\begin{equation}
	\mat{
		\avg{\sx(t)\sxy(0)}_{ss}
		\\
		\avg{\sy(t)\sxy(0)}_{ss}
		\\
		\avg{\sz(t)\sxy(0)}_{ss}
	}
	=
	W
		\mat{
			\avg{\sx\sxy}_{ss}
			\\
			\avg{\sy\sxy}_{ss}
			\\
			\avg{\sz\sxy}_{ss}
		}
		%
	- {\gamma}\avg{\sxy}_{ss}  \vv
	,
	\label{eq::timeDomainSolution}
\end{equation}
with the matrix $W = \Lm^{-1} \parC{\parO{s \id- B}^{-1}}$, and vector $\vv$ given by
\begin{widetext}
\begin{equation}
	W =  \mat{
		f_{2} + \parO{\gm + \gN}f_{1} + \parO{\gm\gN + \Omega^2} f_{0}
		&   \gM f_1 +\gN\gM f_0
		& - \gM \Omega f_0 \\
		\gM f_1 + \gN\gM f_0 
		& f_2 +  \parO{\gN+\gp}f_1 + \gN \gp f_0
		& - \Omega f_1 - \Omega \gp f_0 \\
		 \gM \Omega  f_0
		&   \Omega  f_1 + \Omega \gp f_0
		& f_2 + \gN f_1 + \gNM^2 f_0
	},
\end{equation}
\begin{equation}
	\vv = 
	\Lm^{-1}\parC{ \frac{1}{s}\parO{s \id -B}^{-1} }
	\mat{0 \\0 \\ 1} = \\
	\mat{
		- \gM \Omega f_{-1} \\ - \Omega f_0 - \Omega \gp f_{-1}  \\ f_1 + \gN f_0 + \gNM^2 f_{-1}
	}.
\end{equation}
\end{widetext}
Using properties of the Pauli matrices and Eq.~\eqref{eq::timeDomainSolution},
we can write the correlation function of Eq.~\eqref{eq::corrFunction} as
\begin{equation}
	\avg{ \sp(t) \sm(0)}_{ss}   = K_{0} \erm^{\lr t} + K_{1}\erm^{\lc t} + K_{2} \erm^{\ld t} + K,
	\label{eq:full}
\end{equation}
with
\begin{align}
\begin{split}
	K &= -\frac{\Omega \gamma\parO{\gM+i \gp}}{\lr\lc\ld}\avg{\sm}_{ss},\\
	K_j &= C_j \bigl\{
		\parS{
			2\lambda_j^{2}
			+  \gN \parO{3 \lambda_j + \gN}
			+  \Omega^2
		} \parO{1 + \avg{\sz}_{ss} }\\
        & \qquad +\Omega \parS{\gM+ i \parO{\gp +\lambda_j}}\parO{1 +\frac{\gamma}{\lambda_j}}\avg{\sm}_{ss}
	\bigr\}.
\end{split}
\end{align}
%
%
Finally, putting all of these results together, we obtain from Eq.~\eqref{eq::spectrum} the resonance fluorescence spectrum
%
\begin{equation}
	S(\omega) = \frac{1}{\pi}\sum_{j=0}^{2} \parC{
	\frac{-\parS{
	K_{j}^{R}\lambda_j^{R} + K_{j}^{I} \parO{\omega +\lambda_j^{I}}
			 }
	}{\parO{\lambda_j^{R}}^2 +  \parO{\omega +\lambda_j^{I}}^2}
	} + K\delta(\omega),
	\label{eq::resonanceSpectrum}
\end{equation}
where we have defined real and imaginary parts such that $\lambda_j = \lambda_j^{R} + i \lambda_{j}^{I}$, and $K_j = K_{j}^{R}+i K_{j}^{I}$. 
%
%
Neglecting the last term, we find that the resonance fluorescence spectrum is the sum of three Lorentzians, each multiplied by a function containing corrections that are linear in frequency. Figure \ref{fig:spectra} displays experimental spectra exhibiting such dispersive features.  As we show below, see Eq.~\eqref{eq::LorentzianSpectrum}, in the large Rabi drive limit, the spectrum is purely Lorentzian ($\im{K_i} \sim 0$).
%



\subsection{Reflection spectrum}\label{sec::reflection}
The results of the previous section assume that the above spectrum is directly measured by looking at the fluorescence of the qubit, i.e. ignoring the squeezed bath. However, in the experiment described here, the measurement is performed in reflection implying that the signal fluorescence is combined with radiation from the squeezer reflected by the cavity.   
Following Ref.~\cite{Gardiner1986}, with $u(t) = 0 $ for $t<0$ and $u(t)=1$ for $t>0$, the full reflection spectrum 
from the strongly coupled port (coupling rate $\gamma_{\mathrm{ext}}$)
is related to the time-domain correlation function,

\begin{align}
\begin{split}
	&\avg{\bout^{\dag}(t)  \bout(0)} = \tilde{N} \delta(t)\\
      & \qquad + \gamma_{\mathrm{ext}} u(t) \avg{\commutator{\sp(t)}{\tilde{N} \sm(0)-\tilde{M}\sp(0)}}\\
      & \qquad + \gamma_{\mathrm{ext}} u(-t) \avg{\commutator{\tilde{N} \sp(t)-\tilde{M}\sm(t)}{\sm(0)}}\\
	  & \qquad + \gamma_{\mathrm{ext}} \avg{\sp(t) \sm(0)},
\end{split}
\end{align}
where $\tilde{M}$, $\tilde{N}$ characterize the field in the transmission line.
Hence the reflection spectrum is given by 

\begin{align}\label{eq::spec23}
\begin{split}
	S_{R}(\omega) &=
	\frac{1}{\pi}\int_0^\infty \avg{	\bout^{\dag}(t)  \bout(0)}  {\mathrm{d}}t\\
	& = \frac{\tilde{N}}{2\pi} 
	+ \gamma_{\mathrm{ext}} S(\omega) \\
	%
     + \frac{\gamma_{\mathrm{ext}}}{\pi} 
     &
     \re{ 
     	\int^{\infty}_{0}	\drm t\, 
     	\erm^{i \omega t} 
     	\avg{\commutator{\sp(t)}{\tilde{N} \sm(0)-\tilde{M}\sp(0)}}
     	},
\end{split}
\end{align}
where $S(\omega)$ is the resonance fluorescence spectrum defined in Eq.~\eqref{eq::resonanceSpectrum}. 
Following the standard conventions, the output field operator $\bout$ has units of root frequency and thus the reflection spectrum $S_R(\omega)$ is unitless while $S(\omega)$ has units of inverse frequency.  
Eq.~\eqref{eq::spec23} assumes $M$ real and a single decay channel $\gamma$.
In the more general case where $M = \abs{M}\erm^{2i\Phi} $, and  rewriting the equation in terms of the total damping rate $\gamma = \gamma_{\mathrm{int}}+ \gamma_{\mathrm{ext}}$ and effective field moments $M$ and $N,$
\begin{widetext}
\begin{equation}
	\begin{split}
		S_{R}(\omega) 
		&=
		\frac{N}{2\pi \eta_c} + \gamma  \parO{N+\eta_c} S(\omega) 
		\\ &\quad
		+ \frac{ \gamma}{\pi} \re{\int^{\infty}_{0}	\drm t\, \erm^{i \omega t} 
		 \parS{		\abs{M}\erm^{2 i \Phi}\avg{ \sp(0) \sp(t)} 
					-\abs{M}\erm^{2 i \Phi}\avg{\sp(t) \sp(0)}  
					-N\avg{  \sm(0) \sp(t)}  
			}	}.
	\end{split}
	\label{eq:refSpec}
\end{equation}
\end{widetext}
 
It is important to recall that the above expression is obtained using the broadband squeezing approximation. Since the two-level atom probes the field only in a small frequency range, this approximation is valid for the resonance fluorescence spectrum. The squeezer background is, however, frequency dependent. We take this into account in analyzing the  experimental data by replacing the first term of Eq.~\eqref{eq:refSpec} by the frequency dependent spectrum $N(\omega)$ for the output field of a parametric amplifier~\cite{Collett1984}. With this modification, the analytical expression is found to be in excellent agreement with cascaded master equation simulations (not shown).

In order to relate Eq.~\eqref{eq:refSpec} to the results of the main text, we now examine useful limiting cases. First, we consider the reflection spectrum in the absence of the coherent Rabi drive ($\Omega=0$). In this case, we obtain
(Eq.~\eqref{sqspec} of the main text)
\begin{widetext}
	\begin{align}
		S_R(\omega)
		=\frac{N}{2\pi \eta_c}
		+\frac{1}{2\pi}\frac{\gamma}{2N+1} \left[\frac{\gamma_y(M-(1-\eta_{\mathrm{c}})N)}{\omega^2+\gamma_y^2}
		-\frac{\gamma_x(M+(1-\eta_{\mathrm{c}})N)}{\omega^2+\gamma_x^2}\right],
	\end{align}

\end{widetext}
where, $\gamma_{x,y}=\gamma(N\pm M+1/2)$. From this equation we see that for a lossless single-sided cavity $\eta_c=1$, a fluorescence peak is observed in the reflection spectrum only if $M\neq 0$. The spectrum is a sum of narrow and broad linewidth Lorentzians. However for $\eta_c\neq1$, some of the incident intensity is lost which enhances the relative spectral weight of the negative dip. Finally, in the large Rabi drive limit $\Omega\gg\gamma,\gamma_+,\gamma_-,\gamma_N$, we obtain

    \begin{align}
    \begin{split}
		S_R(\omega)&=\frac{N}{2\pi \eta_c}+\frac{\eta_c}{2\pi}\frac{\gamma\gamma_+}{\omega^2+{\gamma_+}^2} \\
        & \qquad + \frac{1}{8\pi}\frac{\gamma(\gamma_N+\gamma_-)}{(\omega-\Omega)^2+(\gamma_N/2+\gamma_-/2)^2}\\
        & \qquad + \frac{1}{8\pi}\frac{\gamma(\gamma_N+\gamma_-)}{(\omega+\Omega)^2+(\gamma_N/2+\gamma_-/2)^2}
		\label{eq::LorentzianSpectrum}
    \end{split}
  \end{align}
  This equation indicates that for a strong Rabi drive the reflection spectrum is a sum of three Lorentzians centered at $0,\pm\Omega$ and of full widths $2\gamma_+=\gamma(2N+2M\cos2\Phi+1)$ and $\gamma_N+\gamma_-=\gamma(3N-M\cos2\Phi+3/2)$. This explains the cosine dependence with opposite phases of the linewidths as plotted in Fig.~\ref{Mollow_fig} of the main text.

\end{document}